\documentclass[twocolumn,pra,aps,superscriptaddress]{revtex4-1}
\usepackage[T1]{fontenc}
\usepackage[latin9]{inputenc}

\usepackage{amsmath}
\usepackage{amssymb}
\usepackage{xcolor}
\usepackage{bbm}
\usepackage{braket}
\usepackage{mathrsfs}  
\allowdisplaybreaks

\usepackage{graphicx}

\usepackage[colorlinks=true]{hyperref}  
\hypersetup{
    bookmarks=true,         
    unicode=false,          
    pdftoolbar=true,        
    pdfmenubar=true,        
    pdffitwindow=false,     
    pdfstartview={FitH},    
    pdftitle={Unsupervised machine learning and band topology},    
    pdfauthor={Scheurer and Slager},     
    pdfsubject={},   
    pdfcreator={},   
    pdfproducer={}, 
    pdfkeywords={} {} {}, 
    pdfnewwindow=true,      
    colorlinks=true,       
    linkcolor=blue, 
    citecolor=blue,        
    filecolor=magenta,      
    urlcolor=blue,           
	breaklinks=true
}

\definecolor{orange}{rgb}{1,0.5,0}
\newcommand{\equref}[1]{Eq.~(\ref{#1})}

\newcommand{\figref}[1]{Fig.~\ref{#1}}

\newcommand{\refcite}[1]{Ref.~\onlinecite{#1}}

\newcommand{\pdagger}{{\phantom{\dagger}}}
\newcommand{\sect}[1]{\vspace{0.3em}{\it #1.}---}

\newcommand{\be}{\begin{equation}}
\newcommand{\ee}{\end{equation}}
\newcommand{\bea}{\begin{eqnarray}}
\newcommand{\eea}{\end{eqnarray}}

\renewcommand{\vec}[1]{\boldsymbol{#1}}

\newcommand{\appref}[1]{Appendix~\ref{#1}}

\begin{document}
\title{Unsupervised machine learning and band topology}

\author{Mathias S.~Scheurer}
\affiliation{Department of Physics, Harvard University, Cambridge MA 02138, USA}

\author{Robert-Jan Slager}
\affiliation{Department of Physics, Harvard University, Cambridge MA 02138, USA}
\affiliation{TCM Group, Cavendish Laboratory, University of Cambridge, J. J. Thomson Avenue, Cambridge CB3 0HE, United Kingdom}

\date{\today}

\begin{abstract}
The study of topological bandstructures is an active area of research in condensed matter physics and beyond.
Here, we combine recent progress in this field with developments in machine-learning, another rising topic of interest. 
Specifically, we introduce an unsupervised machine-learning approach that searches for and retrieves paths of adiabatic deformations between Hamiltonians, thereby clustering them according to their topological properties. 
The algorithm is general as it does not rely on a specific parameterization of the Hamiltonian and is readily applicable to any symmetry class. We demonstrate the approach using several different models in both one and two spatial dimensions and for different symmetry classes with and without crystalline 
symmetries. Accordingly, it is also shown how trivial and topological phases can be diagnosed upon comparing with a generally designated set of trivial atomic insulators.
\end{abstract}
\maketitle

\sect{Introduction}With the advent of the concept of topological insulators \cite{rmp_qi, rmp_hasan}, considerable research effort 
has focused on further underpinning the theoretical understanding and material realizations of such non-trivial systems. In the past two years, specific progress has been made on systematically categorizing topological band structures upon considering the role of crystal symmetries \cite{Fu_cryst, top1}. The different topological band structures, that is configurations that cannot be mapped into each other without closing the gap and breaking the symmetries under considerations, are obtained as solutions to a combinatorial problem \cite{Us_Prx}, matching the underlying descriptive equivariant K-theory \cite{Us_Prx,Freed_2013}.

Another recently established and very active field of research is concerned with the application of machine learning (ML) techniques to problems in physics \cite{MEHTA20191,RevModPhys.91.045002}. Already in condensed matter physics alone, there have been many different conceptually and practically valuable applications of ML, such as providing variational representations of wavefunctions \cite{Carleo602,CarleoRBMReview}, acceleration of Monte Carlo sampling \cite{PhysRevB.95.041101}, in material science and density functional theory \cite{Schleder_2019,ReviewMaterialScience,osti_1321732,2018arXiv180804733C,2019arXiv191010161C,2019arXiv191203296P,2019arXiv191103580R}, and detection of phase transitions \cite{RogerMelko,WangPCA,PhysRevX.7.031038,QLT,ConfusionScheme,PhysRevB.97.205110,WindingNumberSupervised,WindingNumberSupervised2,PhysRevB.97.115453,PhysRevB.98.174411,PhysRevE.95.062122,OurML,PhysRevB.97.134109,2019arXiv190101963H,ReviewLearningWavefunctions,FruchartLocalMarker,QuantumWalks,2019arXiv190904784T,2019arXiv191010124G,PhysRevLett.122.210503,2019arXiv191010124G,2019arXiv190803469B}. Concerning the latter, it has been established that topological phase transitions are, in general, significantly more difficult to capture than symmetry-breaking phase transitions \cite{PhysRevB.97.045207}, due to the absence of a local order parameter. While some progress has been achieved, most approaches for learning topological phases rely on supervised learning (i.e., require labelled data) and/or manual feature engineering taking into account prior knowledge of the phases. 
In \refcite{OurML}, however, an unsupervised ML approach has been established that can classify samples based on global topological properties from raw data, i.e., without any feature engineering. The key idea is to view the samples as nodes on a graph with connections defined by the local similarity, $K_{l,l'}$, of pairs $l$, $l'$ of samples. The global structure of the graph can be represented in a low-dimensional embedding constructed from diffusion maps \cite{Coifman7426}; this, in turn, reveals the distinct topological equivalence classes in the data. 

\begin{figure}[b]
   \centering
    \includegraphics[width=0.95\linewidth]{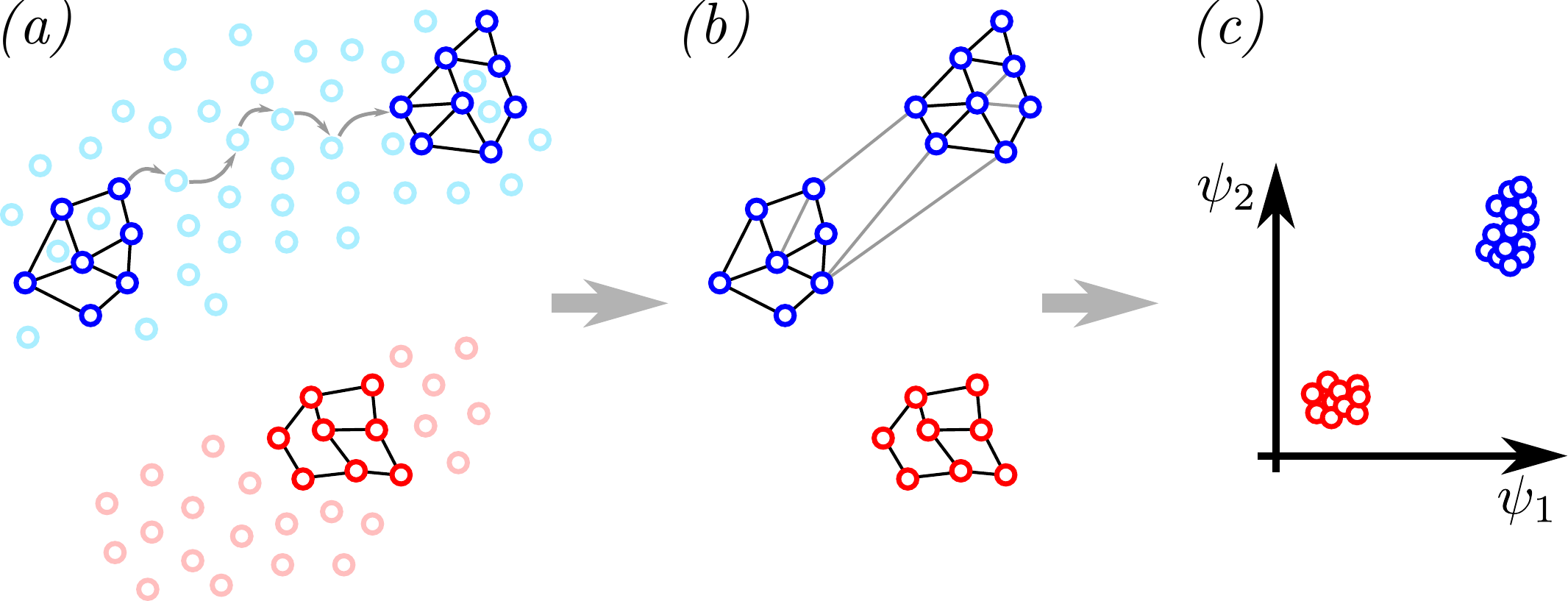}
    \caption{
    Instead of randomly sampling Hamiltonians of the symmetry class of interest [light blue dots in (a)], a path finding algorithm is employed to retrieve adiabatic deformations (gray arrows) between Hamiltonians (dark red/blue dots). 
    This allows to construct an effective graph, see (b), which we use to find a low-dimensional embedding (c) that reveals the topologically distinct classes.}
    \label{fig:Schematics}
\end{figure}

When detecting \textit{bulk topological order} \cite{SubirReview}, e.g., of the two-dimensional (2D) XY model, this procedure can be directly applied to (Monte Carlo) snapshots obtained for these systems \cite{OurML}, revealing the presence or absence of superselection sectors \cite{KitaevSuperselectionSectors}. In this paper, we are interested in \textit{band topology}. In that case, we are typically only given a few Hamiltonians, which are our ``samples'' in the ML context, represented by the dark blue and red dots in \figref{fig:Schematics}(a), that we want to cluster according to whether they can be deformed into each other or not. These, however, do not a priori cover the entire space of Hamiltonians of that symmetry class. To make sure we are not missing any paths between Hamiltonians, we could just randomly sample Hamiltonians of that symmetry class (light dots), which has very recently been suggested in the context of a different ML setup \cite{2019arXiv190803469B}, and subsequently apply the procedure of \refcite{OurML} to this larger set of samples. Since the required amount of random samples grows rapidly with the size of the Hilbert space, we here propose a different ML procedure that  systematically retrieves  adiabatic paths between Hamiltonians (little gray arrows). In this way, we construct an effective graph, see \figref{fig:Schematics}(b), that we take as starting point yielding a low-dimensional embedding, (c), that reflects the topological sectors. In this space, we then apply conventional $k$-means clustering.

\sect{Algorithm}We now present the general ML algorithm.  
Suppose we are given a set of $m$ Hamiltonians with $N_b$ bands, $\{h^{l}_{\vec{k}} \in \mathbbm{C}^{N_b\times N_b},l=1,\dots ,m\}$, where we have assumed that the system is translationally invariant and, hence, momentum $\vec{k}$ is a good quantum number. Our goal is to classify the ``samples'' $l=1,\dots ,m$ topologically, i.e., decide which Hamiltonians can be deformed into each other without closing the gap or breaking a certain set of symmetries. The latter can include time-reversal symmetry (TRS), $\Theta$, particle-hole symmetry (PHS), $\Xi$, and chiral symmetry, $C$, as in the usual Altland-Zirnbauer (AZ) classification, but also any other set of unitary symmetries constituting a group $\mathscr{G}$. 

To define a notion of what classifies as a ``continuous deformation'', we introduce the following \textit{measure of $\vec{k}$-local similarity} between Hamiltonians $l$ and $l'$ ($N_k$ is the number of $\vec{k}$ points in the sum)
\begin{equation}
    S_{l,l'} = \frac{1}{N_k} \sum_{\vec{k}}  \frac{1}{N_b}\sum_{n=1}^{N_b} \left|\braket{\psi_{n\vec{k}}^l|\psi_{n\vec{k}}^{l'}} \right|^2, \label{MeasureOfSimilarity}
\end{equation}
with eigenstates, $\ket{\psi^l_{n\vec{k}}}$, satisfying $h^l_{\vec{k}}\ket{\psi^l_{n\vec{k}}} = \epsilon^l_{n\vec{k}}\ket{\psi^l_{n\vec{k}}}$, $\epsilon^l_{n\vec{k}} < \epsilon^l_{n'\vec{k}}$ for $n<n'$ (for now, we neglect degeneracies). The normalization in \equref{MeasureOfSimilarity} is chosen such that $S_{l,l}=1$. It further holds $S_{l,l'}=S_{l',l}$, $0\leq S_{l,l'} \leq 1$, and $S_{l,l'}=1$ if and only if $h^{l}_{\vec{k}}$ and $h^{l'}_{\vec{k}}$ are identical except for a deformation of the band energies, that does not close any gap in the system. 
If the set of Hamiltonians already contains all relevant adiabatic paths, we can simply directly use the approach of \refcite{OurML} with connections $ K^{0}_{l,l'}=\exp(-(1-S_{l,l'})/\epsilon)$ to perform the topological analysis, where $\epsilon$ is a suitably chosen coarse-graining parameter. As discussed above, this is however very unlikely. To overcome this issue, we will make use of the fact that the spectra of two Hamiltonians $h_{\vec{k}}$ and $h'_{\vec{k}}$ are topologically equivalent if $h'_{\vec{k}}=U^\pdagger_{\vec{k}}h^\pdagger_{\vec{k}} U^\dagger_{\vec{k}}$ with unitary $U_{\vec{k}} = e^{i\varphi_{\vec{k}}\Lambda}$ that respects the symmetries of the symmetry class under consideration; here $\Lambda$ is the associated generator and $\varphi_{\vec{k}}$ is required to be a continuous function of $\vec{k}$. On the level of states, this ``adiabatic deformation'' corresponds to $\ket{\psi_{n\vec{k}}} \rightarrow  U_{\vec{k}} \ket{\psi_{n\vec{k}}}$. For each pair of Hamiltonians $l$ and $l'$ in \equref{MeasureOfSimilarity}, we perform a sequence of deformations, $\ket{\psi^{l'}_{n\vec{k}}} \rightarrow \ket{\psi^{l',1}_{n\vec{k}}} \rightarrow \ket{\psi^{l',2}_{n\vec{k}}} \rightarrow \dots \rightarrow \ket{\psi^{l',N_f}_{n\vec{k}}}$, to maximize the value of $S_{l,l'}$. The resultant final value $S^f_{l,l'}$ of the similarity measure, i.e., $S_{l,l'}$ with $\ket{\psi^{l'}_{n\vec{k}}}$ replaced by $\ket{\psi^{l',N_f}_{n\vec{k}}}$, will be used as input for $K_{l,l'}=\exp(-(1-S^f_{l,l'})/\epsilon)$. 

Before discussing our ML procedure to find this path, we will come back to the issue of degeneracies and further refine the form of $S_{l,l'}$ in \equref{MeasureOfSimilarity}, allowing for the possibility that we might only be interested in keeping a specific subset of the gaps of the system open while all others are permitted to close.
Taking such different partitions of bands has recently been linked to new forms of topology and are therefore interesting in their own right \cite{bouhon2019nonabelian,fragile, Us_Wilson}.
To this end, we replace the kernel in \equref{MeasureOfSimilarity} by the refined similarity measure
\begin{equation}
    S_{l,l'} = \frac{1}{N_k} \sum_{\vec{k}}  \frac{1}{\widetilde{N}_b}\sum_{n=1}^{N_{b}} \sum_{n'\in \mathcal{S}_n} \left|\braket{\psi_{n\vec{k}}^l|\psi_{n'\vec{k}}^{l'}} \right|^2, \label{RefinedSimilarityMeasure}
\end{equation}
where $\mathcal{S}_n$ is the subset of bands $\{1,\dots,N_b\}$ that band $n$ is allowed to close gap or is degenerate with. Further, $\widetilde{N}_b=\sum_{n=1}^{N_b} |\mathcal{S}_n|$ with $|\mathcal{S}_n|$ denoting the number of elements in $\mathcal{S}_n$. Our previous form in \equref{MeasureOfSimilarity} corresponds to $\mathcal{S}_n=\{n\}$ with $|\mathcal{S}_n|=1$. To illustrate this in an explicit context, consider the scenario of four bands, $n=1,2,3,4$, assuming that we are only interested in the topological properties of the half-filled gap, i.e., between $n=2$ and $n=3$, while the other gaps between $n=1$, $n=2$ and between $n=3$, $n=4$ can be closed (or are zero due to a degeneracy). In that case, we have $S_{1}=S_{2} = \{1,2\}$ and $S_{3}=S_{4} = \{3,4\}$.

Next, we discuss the approach we use here to find the path between two Hamiltonians $l$ and $l'$. 
To construct the unitary transformation, $U_{\vec{k}} = e^{i\varphi_{\vec{k}}\Lambda}$, of the Markov chain of deformations of Hamiltonian $l'$, we first randomly sample a Hermitian generator $\Lambda \in \mathbbm{C}^{N_b\times N_b}$. 
We find the optimal momentum dependence of $\varphi_{\vec{k}}$ by expanding the change of the measure of similarity in \equref{RefinedSimilarityMeasure} under $\ket{\psi_{n\vec{k}}^{l'}} \rightarrow U_{\vec{k}}\ket{\psi_{n\vec{k}}^{l'}}$ in $\varphi_{\vec{k}}$ to derive the gradient ascent expression
\begin{equation}
    \varphi_{\vec{k}} =  - \eta\frac{1}{\widetilde{N}_b}\sum_{n=1}^{N_b} \sum_{n'\in \mathcal{S}_n} \hspace{-0.3em} \text{Im}\left[\braket{\psi_{n\vec{k}}^l|\Lambda|\psi_{n'\vec{k}}^{l'}}\braket{\psi_{n'\vec{k}}^{l'}|\psi_{n\vec{k}}^{l}} \right] \label{VarphiFormGD}
\end{equation}
with learning rate $\eta \in \mathbb{R}^+$. To ensure that the ``deformation'' performed on the Hamiltonian is smooth in momentum space, we only accept the update if the maximum gradient of $\varphi_{\vec{k}}$ in momentum space is smaller than the cutoff $\Lambda_{\varphi}$. We also compute the overall smoothness of the states, $\text{min}_j \sum_{n} \sum_{n'\in \mathcal{S}_n}  | \braket{\psi_{n\vec{k}_j}^{l'}|\psi_{n'\vec{k}_{j+1}}^{l'}}|^2/N_b$, where $\vec{k}_j$ are the momentum grid points and periodicity of the Brillouin zone is taken into account. If it drops below a cutoff value, $\Lambda_s$, during the deformation Markov chain, we abort and set $S_{l,l'}=0$.

Besides smoothness of transformations, we also need to make sure that the deformation parametrized by $\Lambda$ and $\varphi_{\vec{k}}$ preserves the symmetries we are interested in. Let us first focus on unitary symmetries and denote the representation of $g\in \mathscr{G}$ in momentum and in the $N_b$-dimensional space of the Hamiltonian by $\mathcal{R}_v(g)$ and $\mathcal{R}_\psi(g)$, respectively. The resulting constraint $\varphi_{\mathcal{R}^{-1}_v(g)\vec{k}}\mathcal{R}^\pdagger_\psi(g) \Lambda  \mathcal{R}^\dagger_\psi(g) = \varphi_{\vec{k}} \Lambda$ can be imposed by explicit symmetrization,
\begin{equation}
    \Lambda \varphi_{\vec{k}}\quad \longrightarrow \quad \frac{1}{|\mathscr{G}|} \sum_{g\in\mathscr{G}} \varphi_{\mathcal{R}^{-1}_v(g)\vec{k}} \mathcal{R}^\pdagger_\psi(g) \Lambda  \mathcal{R}^\dagger_\psi(g) , \label{ExplicitSymmetrization}
\end{equation}
where $|\mathscr{G}|$ denotes the number of elements of $\mathscr{G}$ \footnote{As we show in \appref{SteepestAscent}, this symmetrization is consistent with the steepest ascent condition used to derive \equref{VarphiFormGD}.}.

\begin{figure}[t]
   \centering
    \includegraphics[width=\linewidth]{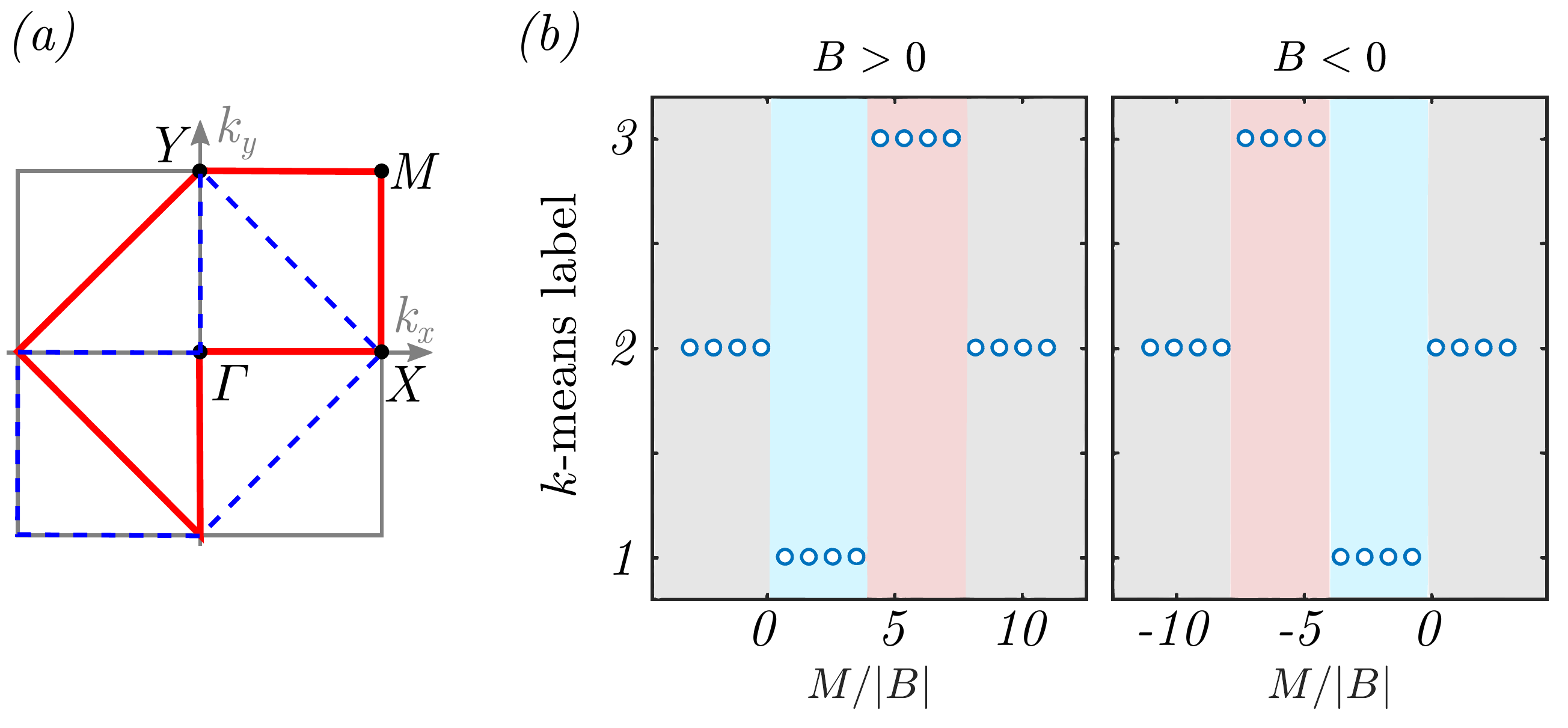}
    \caption{
    (a) 1D closed path in the Brillouin zone (red solid) going through the indicated high-symmetry points and its time-reversal partner (blue dashed) that, due to symmetry, is effectively also taken into account. (b) $k$-means labels for the BHZ model (\ref{SimpleBHZHamiltonian}) as a function of $M$ for $B>0$ (left panel) and $B<0$ (right panel). Hyperparameters: $N_k=180$,  $\epsilon=0.05$, $\eta=0.4$, $\Lambda_s=0.85$, $\Lambda_\varphi=0.1$, $N_{\text{kick}}=50$, $f=0.5$.}
    \label{fig:BHZModel}
\end{figure}

We distinguish two different ways of implementing this procedure in practice:
if we sum over (a discretized version of) the full Brillouin zone in \equref{RefinedSimilarityMeasure} and compute $\varphi_{\vec{k}}$ for all $\vec{k}$, we can explicitly perform the symmetrization in \equref{ExplicitSymmetrization}. This is what we do in the examples in one-dimension (1D) below. In higher dimensions, we can (but do not have to) speed up the algorithm by taking into account that the topological properties are encoded in the behavior, i.e the existence of windings, of the Wilson operators \cite{Us_Wilson, Wi1, Wi2} along 1D cuts in the Brillouin zone that go through all high-symmetry points. Therefore, we can restrict the momenta $\vec{k}$ in Eqs.~(\ref{MeasureOfSimilarity}-\ref{VarphiFormGD}) to these cuts. Furthermore, if the diagonalization of the Hamiltonian is the most computationally expensive step, we can use the symmetries of the system to restrict this path to symmetry-inequivalent momenta only [see, e.g., \figref{fig:BHZModel}(a)]. To compute $\varphi_{\mathcal{R}^{-1}_v(g)\vec{k}}$ in \equref{ExplicitSymmetrization} in that case, we use that the symmetry of the Hamiltonian implies that $\varphi_{\mathcal{R}^{-1}_v(g)\vec{k}}$ is given by \equref{VarphiFormGD} with $\Lambda$ replaced by $\mathcal{R}^\pdagger_\psi(g)\Lambda\mathcal{R}^\dagger_\psi(g)$ on the right-hand side.

Finally, we discuss how to take into account the symmetries of the conventional AZ classes. Imposing chiral symmetry, $C h_{\vec{k}} C^\dagger = -h_{\vec{k}}$, is very straightforward as it simply amounts to symmetrizing the generator $\Lambda \rightarrow (\Lambda + C\Lambda C^\dagger)/2$ right after sampling it. It is, thus, only left to analyze the case of only one of TRS and PHS. Focusing for notational simplicity on the former (the exact same applies to PHS), the associated constraint $\varphi_{\vec{k}} \Theta \Lambda  \Theta^\dagger = -\varphi_{-\vec{k}} \Lambda$ is rectified by replacing $\varphi_{\vec{k}} \rightarrow (\varphi_{\vec{k}} \Lambda - \varphi_{-\vec{k}} \Theta\Lambda\Theta^\dagger)/2$. If we want to restrict the momentum points to an irreducible set, we do not have to compute both $\varphi_{\vec{k}}$ and $\varphi_{-\vec{k}}$: using the TRS of the system, the symmetrization can also be restated as $\varphi_{\vec{k}} \rightarrow \sum_{p=\pm} \varphi^p_{\vec{k}} \Lambda^p/2$, where $\Lambda^{\pm} = (\Lambda \pm \Theta \Lambda \Theta^\dagger)/2$ and $\varphi_{\vec{k}}^{\pm}$ are given by \equref{VarphiFormGD} with $\Lambda$ replaced by $\Lambda^{\pm}$.

We close the general discussion of the algorithm 
with a few remarks about its implementation. For specific Hamiltonians with many additional symmetries (beyond the symmetry group we are interested in), the procedure becomes more reliable when performing initial ``kicks'' to it: unless the two samples $l$ and $l'$ are already close to each other according to the measure in \equref{RefinedSimilarityMeasure}, we sample a few random, $\vec{k}$-independent unitary transformations $U_j$, $j=1,2,\dots N_{\text{kick}}$ (properly symmetrized), and take the one that leads to the largest value of \equref{RefinedSimilarityMeasure} with $\ket{\psi_{n'\vec{k}}^{l'}}\rightarrow U_j \ket{\psi_{n'\vec{k}}^{l'}}$. 
Similarly, we also noted that the iteration converges significantly faster if every other gradient ascent step is replaced by just performing a $\vec{k}$-independent update: we sample $\varphi$ from a Gaussian distribution and generate a symmetrized generator $\Lambda$. Only if \equref{RefinedSimilarityMeasure} with $\ket{\psi_{n'\vec{k}}^{l'}}\rightarrow e^{i\varphi \Lambda} \ket{\psi_{n'\vec{k}}^{l'}}$ is larger than before, we accept the update. Finally, note that our approach does not require that all paths are identified perfectly and we do not have to perform the path-finding iteration for all combinations of $l$ and $l'$. Due to the stability of the approach outlined above to perturbations, it also works just as fine if we only iterate for a randomly chosen fraction $f \leq 1$ of pairs of samples, reducing the computational cost of the algorithm. 

\begin{figure}
   \centering 
    \includegraphics[width=\linewidth]{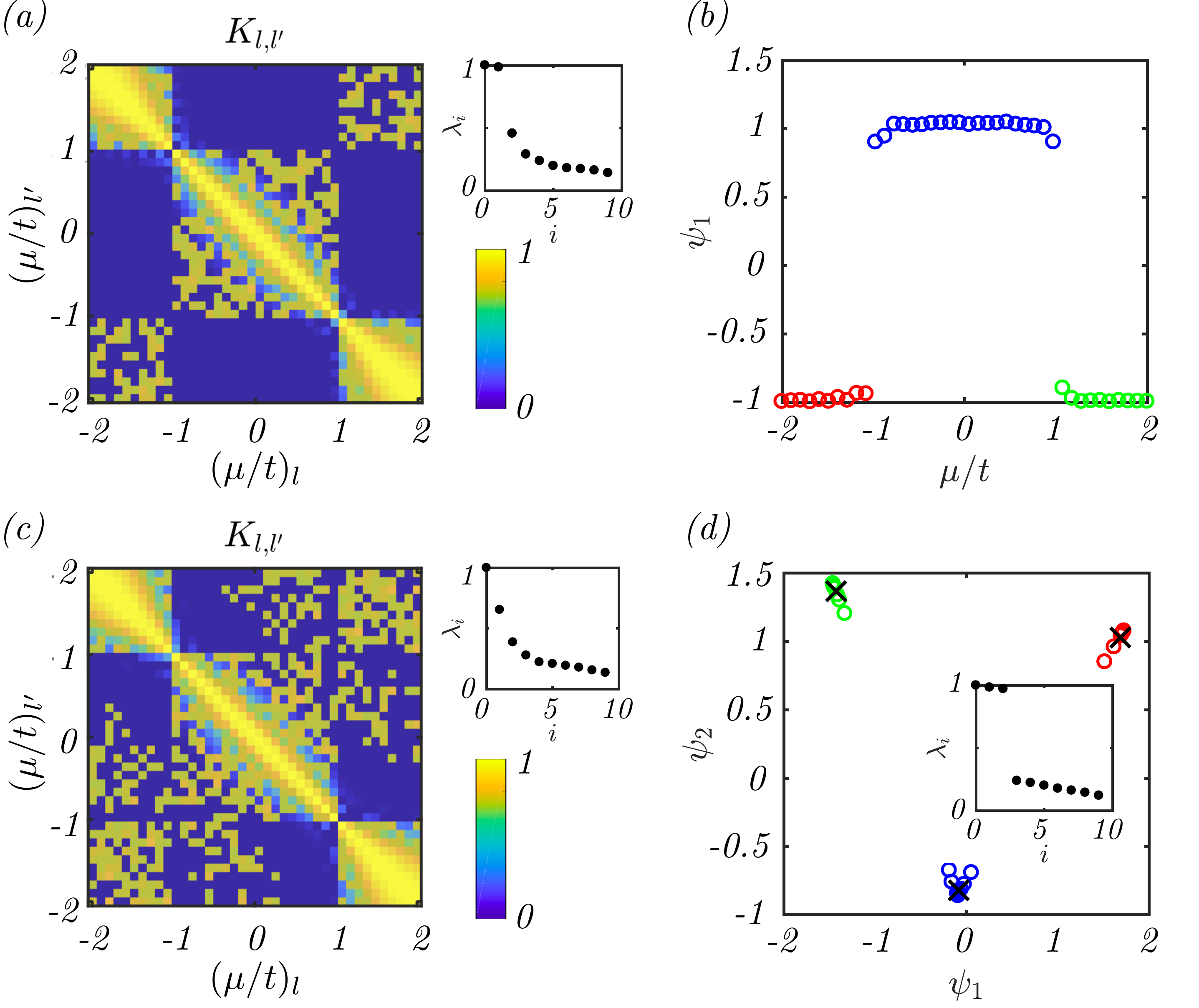}
    \caption{
    (a) The kernel and resulting eigenvalues (inset) for $m=40$ samples of Hamiltonians (\ref{KitaevHam}) embedded in class AIII, together with (b) the dominant component, $\psi_1$, as a function of $\mu/t$. (c) is the same as (a) but for class A. As can be seen in (d), for class BDI, we have $N_t=2$ dominant eigenvectors (inset) that reveal three clusters (main panel), where colors refer to values of $\mu/t$ as in (b) and the crosses to the $k$-means centroids. Hyperparameters: $N_k=50$,  $\epsilon=0.03$, $\eta=0.3$, $\Lambda_s=0.5$, $\Lambda_\varphi=0.1$, $N_{\text{kick}}=30$, $f=0.5$.}
    \label{fig:AZIn1D}
\end{figure} 
\sect{Altland-Zirnbauer in 1D}In the remainder, we apply our proposed ML scheme to a variety of different models and symmetry classes. 
To start as simple as possible, let us consider the conventional AZ classes in 1D and $N_b=2$-band models. The set of Hamiltonians we want to cluster according to topological properties are taken to be of the form of the Kitaev model \cite{Kitaev_2001},   
\begin{equation}
h_k = \Delta \sin k\, \sigma_1 + (-t \cos k - \mu) \sigma_3. \label{KitaevHam}
\end{equation}
Here $\sigma_j$ are Pauli matrices and we will set $\Delta=t$ for concreteness such that we are left with only one dimensionless parameter $\mu/t$ to parametrize the phase diagram. For topological classification, there are different ``ensembles'' of Hamiltonians we can embed \equref{KitaevHam} in. 

To begin with class AIII, we only impose chiral symmetry $C=i\sigma_2$. In \figref{fig:AZIn1D}(a), we illustrate the resulting kernel associated with the effective graph after searching for additional deformations between the Hamiltonians. 
We see that the ML procedure correctly identifies that there are adiabatic paths between Hamiltonians with $\mu/t<-1$ and $\mu/t>1$; note that these paths are not present in the data set of Hamiltonians as the direct overlap according to the similarity measure (\ref{MeasureOfSimilarity}) is very small between Hamiltonians in \equref{KitaevHam} with $\mu/t<-1$ and $\mu/t>1$. We now take this kernel as input for the ML procedure of \refcite{OurML}, which yields a set of eigenvalues, $\lambda_i$, and associated eigenvectors, $(\psi_i)_l$, $i=0,1,\dots$. The number $N_t$ of $\lambda_i$ (exponentially) close to one is equal to the number of distinct topological equivalence classes in the data; as expected, we here get $N_t=2$, see inset in \figref{fig:AZIn1D}(a), with a clear gap to the subleading eigenvalues. Furthermore, the clustering according to topological features is visible in the low-dimensional embedding $l \rightarrow [(\psi_1)_l,\dots,(\psi_{N_t-1})_l]$. Consequently, in our case here, $\psi_1$ should be sufficient and, indeed, we see in \figref{fig:AZIn1D}(b) that $\psi_1$ correctly identifies the two topologically distinct regimes $|\mu/t|>1$ and $|\mu/t|<1$.

\begin{figure}
   \centering
    \includegraphics[width=\linewidth]{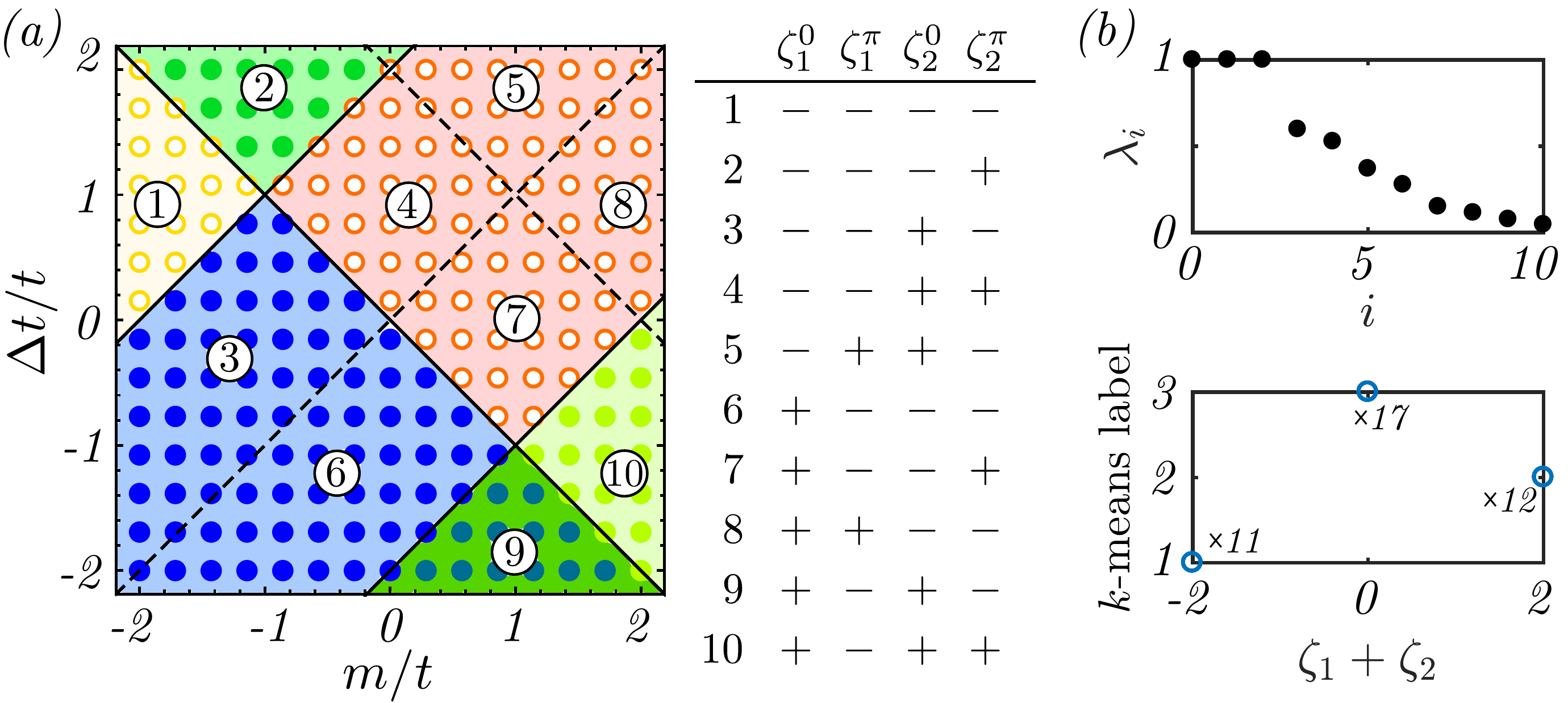}
    \caption{
    The different colors of the circles in (a) correspond to different $k$-means labels obtained for the $m=15 \times 14$ samples of Hamiltonians in \equref{4bandInvModel} from the low-dimensional embedding of our ML procedure ($N_t=6$ is found). The table (right panel) shows the eigenvalues of $P$ in the respective parts of the phase diagram. Filled (empty) circles correspond to Hamltonians not connected (connected) to a trivial atomic insulator. Sampling $m=40$ trivial insulators, we identify three different classes (see eigenvalues upper panel) correlated with $\zeta_1 + \zeta_2$ (scatterplot $k$-means vs.~$\zeta_1 + \zeta_2$, lower panel). We use $N_k=50$,  $\epsilon=0.03$, $\eta=0.8$, $\Lambda_s=0.5$, $\Lambda_\varphi=0.1$, $N_{\text{kick}}=0$, $f=0.2$ (a), $f=0.5$ (b).}
    \label{fig:CrystallineSymmetries}
\end{figure}

Next, we consider class A, i.e., we do not impose any symmetry at all. As in any odd dimension, all phases must be equivalent and, indeed, the ML procedure finds paths between all three regimes $\mu/t < -1$, $-1<\mu/t<1$, and $\mu/t>1$, see \figref{fig:AZIn1D}(c), and there is only one dominant eigenvalue (see inset).

Finally, let us embed \equref{KitaevHam} in class BDI, i.e., impose both TRS, $\Theta=i\sigma_3\mathcal{K}$, with complex conjugation $\mathcal{K}$, and PHS, $\Xi=i\sigma_1\mathcal{K}$ (automatically implying chiral symmetry). As can be seen in \figref{fig:AZIn1D}(d), we now find three quasi-degenerate dominant eigenvalues and three clusters in the associated low-dimensional embedding, $[\psi_1,\psi_2]$, corresponding to the three phases separated by gap closings at $\mu/t=\pm 1$. Although the stable topological $\mathbb{Z}$-invariant of class BDI is the same for $\mu/t<-1$ and $\mu/t>1$, this is, indeed, the correct answer within the subspace of 2-band models (see \appref{ExplicitDeformations}). To investigate this further, we have taken two identical copies of \equref{KitaevHam}, yielding a $N_b=4$-band model; in that case, the ML yields two sectors (not shown) corresponding $|\mu/t|>1$ and $|\mu/t|<1$, as anticipated from a theoretical point of view. 

\sect{Crystalline symmetries}To illustrate the application to topological phases stabilized by crystalline symmetries, let us consider the $N_b=4$-band model
\begin{align}\begin{split}
    &h_k =  \sin k \,(a_1 \sigma_3\otimes\tau_1 + a_2 \,\sigma_0\otimes\tau_2) + [t(1-\cos k)\\ &\,\,-m]\sigma_0\otimes\tau_3 + \Delta t \cos k \, \sigma_0 \otimes (\tau_0+\tau_3) + \delta \sigma_1 \otimes \tau_3, \label{4bandInvModel}
\end{split}\end{align}
where both $\sigma_j$ and $\tau_j$ are Pauli matrices acting in different spaces. This model exhibits inversion symmetry, $P h_{-k} P^\dagger = h_k$ with $P=\sigma_0 \otimes \tau_3$, which is the only symmetry we will impose. For concreteness, we will focus on $a_1=a_2=t$ and $\delta/t=2$ in the following. Allowing for gap closings only between the pairs of bands $n=1,2$ and $n=3,4$, the phase diagram obtained by our ML approach is shown in \figref{fig:CrystallineSymmetries}(a), left panel. We see by comparison with the eigenvalues $\zeta_n^{k_\Theta}$ of $P$ at the $\Theta$-invariant momenta $k_\Theta=0,\pi$ (right panel), that it correctly identifies that $\zeta_1^{0}+\zeta_2^{0}$ and $\zeta_1^{\pi}+\zeta_2^{\pi}$ characterize the distinct topological equivalence classes of bandstructures \cite{PhysRevB.83.245132}.

\sect{Referencing}Notable recent schemes essentially indicate crystalline band topology \cite{Po_tci,Bradlyn_tci} by using the constraints of \refcite{Us_Prx} and comparing to a trivial reference subset, e.g., band structures that can be obtained after Fourier transforming localized real-space Wannier states. Our scheme is flexible to allow for deformations up to an arbitrary reference. To exemplify this, we consider this for the 1D model with inversion symmetry in \equref{4bandInvModel}. 
Note, however, that 1D is special and there is no unique choice for this trivial subset; in analogy to the SSH chain, we here choose states without a filling anomaly to be trivial \cite{Hwang_2019}, which amount to taking the set of momentum-independent, inversion symmetric Hamiltonians as reference (see \appref{ReferencingDetails}). As a first step, we randomly sample momentum-independent Hamiltonians with inversion symmetry and apply our ML procedure to classify them. As can be seen in \figref{fig:CrystallineSymmetries}(b), we find the correct number of three different equivalence classes, which correspond to the net parity $\zeta_1 + \zeta_2$, with $\zeta_n\equiv \zeta_n(0)=\zeta_n(\pi)$. As a second step, we take one member of each of those three classes of trivial states and add them to set of Hamiltonians of \figref{fig:CrystallineSymmetries}(a). This allows us to determine which of the phases in the phase diagram are adiabtically connected to a trivial reference set (open circles) and, thus, trivial, and which are not (solid circles, topological).

\sect{Two dimensions}To illustrate how our approach can be applied in higher dimensions, we next discuss a paradigmatic model of a 2D topological insulator, the BHZ model \cite{Bernevig1757}. Its Hamiltonian reads as
\begin{equation}
h_{\vec{k}}= a_1 \sin(k_x)\,\Gamma_x+a_2\sin(k_y)\,\Gamma_y+M_{\vec{k}}\,\Gamma_0, \label{SimpleBHZHamiltonian}
\end{equation}
where $M_{\vec{k}}=M-2B[2-\cos(k_x)-\cos(k_y)]$, $\Gamma_x=\sigma_3\otimes\tau_1$, $\Gamma_y=\sigma_0\otimes\tau_2$, and $\Gamma_0=\sigma_0\otimes\tau_3$. We will here focus on $a_1=a_2=|B|$ and only impose TRS with $\Theta = i\sigma_2 \otimes \tau_0 \mathcal{K}$ as symmetry. As already discussed in more detail above, we do not have to take into account all 2D momenta, but can, instead, focus on 1D, closed paths in the Brillouin zone that go through all high-symmetry momenta; for the BHZ model with TRS, the latter are $\Gamma$, $X$, $Y$, and $M$ \cite{Us_Prx}, and we will take the path shown in red in \figref{fig:BHZModel}(a). We have checked that our results are not altered when choosing a different path that includes the high-symmetry points. To test our ML approach, we take BHZ Hamiltonians with different values of $M/|B|$ for both signs of $B$ as input. As can be seen in \figref{fig:BHZModel}(b), it finds the nontrivial paths between Hamiltonians with opposite signs of $B$ and correctly identifies the three topologically distinct phases \cite{Skyrmions2d}.

\sect{Conclusion}In this work, we have presented and demonstrated a ML algorithm that identifies adiabatic paths between Hamiltonians, readily applicable to any arbitrary symmetry class, and thereby allows to construct topological phase diagrams without supervision. We hope that future work based on reinforcement or in combination with supervised learning can further improve our approach. Due to the flexibility of the algorithm, which
, in principle, can also be applied to interacting systems, we believe that it will help deepen our understanding of topological band theory.

\begin{acknowledgments}
In particular, we thank Joaquin Rodriguez-Nieva for an inspiring previous collaboration \cite{OurML} and for useful comments on this work. We further thank  Henry Shackleton for useful discussions.
M.S.S acknowledges support from the National Science Foundation under Grant No.~DMR-1664842. R.~J.~S appreciatively acknowledges funding via Ashvin Vishwanath at Harvard University, Trinity college of the University of Cambridge, 
and the Winton programme at the University of Cambridge. 
\end{acknowledgments}

%


\begin{appendix}
\section{Explicit deformations of 1D models}
\label{ExplicitDeformations}
To understand, why the phases of the Kitaev model (\ref{KitaevHam}) with $\mu/t>1$ and $\mu/t<-1$  cannot be deformed into each other within class BDI, let us write $U_k = e^{i\sum_{j=1}^3\varphi_k^j \sigma_j}$. The chiral symmetry $C=\Theta\Xi=i\sigma_2$ implies $\varphi_k^1 = \varphi_k^3 =0$ and, hence,
\begin{equation}
	U_k = e^{i\varphi_k \sigma_2}. \label{ChiralTrafo}
\end{equation}
Parametrizing the Hamiltonian in \equref{KitaevHam} as $h_k = \vec{g}_k \cdot (\sigma_1,\sigma_3)$, $\vec{g}_k = (g_1(k),g_2(k))$, it is clear that \equref{ChiralTrafo} just corresponds to a momentum-dependent rotation of the unit vector $\hat{\vec{g}}_k:=\vec{g}_k/|\vec{g}_k|$. TRS and PHS lead to $\varphi_k = -\varphi_{-k}$ in \equref{ChiralTrafo}: the rotational angle has to be an odd function of $k$ and can, thus, never rotate $\hat{\vec{g}}_{\vec{k}} \sim (0,-1)$ into $\hat{\vec{g}}_{\vec{k}} \sim (0,1)$, which correspond to $\mu/t\rightarrow \infty$ and $\mu/t\rightarrow -\infty$, respectively. This is why the ML classifies $\mu/t < -1$, $\mu/t > 1$, and, of course, $-1 < \mu/t <1$, as three topologically distinct phases, see \figref{fig:AZIn1D}(d). 

When we remove TRS and PHS and only keep $C$, which correspond to class AIII, the allowed unitary transformations are just given by \equref{ChiralTrafo} without further constraints on the momentum dependence of $k$. Therefore, it becomes possible to ``deform'' the phases with large positive and large negative $\mu$ into each other.

The difference of the phases with  $|\mu/t|>1$ also disappears within class BDI when we double the size of the Hilbert space and consider two copies of the Kitaev model in \equref{KitaevHam},
\begin{equation}
h^{(4)}_k = \sum_j \alpha_j \left[ \Delta \sin k\, \sigma_1 + (-t \cos k - \mu) \sigma_3 \right] \otimes P_j. \label{KitaevHamExt}
\end{equation}
Here $P_1 = (\tau_0 + \tau_z)/2$ and $P_2 = (\tau_0 - \tau_z)/2$ are projectors and $\alpha_1 < \alpha_2 \in \mathbbm{R}^+$ to ensure singly-degenerate bands. We will choose the  extensions of time-reversal, $\Theta = \sigma_3 \otimes \tau_0 \mathcal{K}$, and of particle-hole symmetry, $\Xi = \sigma_1 \otimes \tau_0 \mathcal{K}$, which are trivial in the new internal, ``orbital'' degree of freedom described by the Pauli matrices $\tau_j$; clearly, $h^{(4)}_k$ respects both of these symmetries.

We can analytically understand that $\mu/t > 1$ and $\mu/t<-1$ are now adiabatically connected and, hence, topologically equivalent as  found by our ML approach. To see this, let us take the limit $|\mu|\rightarrow \infty$, for which $h^{(4)}_k \sim -\mu \sigma_3 \otimes (\alpha_+ \tau_0+\alpha_- \tau_3)$ wih $\alpha_p = (\alpha_1 + p\,\alpha_2)/2$. We can now define the one-parameter family of Hamiltonians $h^{(4)}_k(\varphi) = U^\dagger_{\varphi}h^{(4)}_k U^\pdagger_{\varphi}$, $U_{\varphi} = \exp(i \varphi (\sigma_2+\sigma_0)\otimes \tau_2)$ which, by tuning $\varphi$ from $0$ to $\pi/2$,  interpolates between positive and negative $\mu$ without closing the gap, while preserving PHS and TRS.

\section{Referencing in 1D and beyond}
\label{ReferencingDetails}
Recent schemes to indicate band topology \cite{Po_tci,Bradlyn_tci} in essence profit from general combinatorial constraints that determine different classes that cannot adiabatically  be  deformed into each other \cite{Us_Prx}, i.e without closing the gap, and subsequently determine which of those have an atomic limit. In other words, these approaches compare to a reference that is defined as the trivial subset, which in this case amounts to all band structures that can be obtained after Fourier transforming real space Wannier states. As shown in the main text, our scheme is flexible to allow for such deformations up to an arbitrary reference state as well as to different partitionings of bands, meaning that we can allow for the closure of some gaps and determine the topology relative to the other gaps. We note that the recently found fragile topologies \cite{fragile, Us_Wilson} are an ultimate consequence of considering topological structures of a different partitioning of the bands rather than the usual valence and conduction set. Indeed, these notions are only stable upon excluding, that is not taking into account, certain additional trivial sets of bands. 

We can make the referencing also concrete for the 1D case. However, due to the peculiarities of one spatial dimension, one needs to exercise some caution as in this case insulators are either atomic or obstructed atomic \cite{Hwang_2019, obs1, obs2}. In case for the inversion symmetric model, for example, both regimes correspond to an occupied Wyckhoff position, either A or B. However, only the latter B possibility gives a different parity eigenvalue in momentum space after Fourier transforming with regard to the $k=0$ and $k=\pi$ momenta. Cutting the system in real space in this case then requires the addition or removal of charge, resulting in a filling anomaly \cite{Anomaly}. It is therefore reasonable to define the complement configurations, that is those corresponding to the A sites having no filling anomaly, as the ``trivial'' reference. Our approach then correctly captures the classes with respect to this state, which in some cases can even be inspected using simple static Hamiltonians. Specifically, to sample the trivial insulators for \figref{fig:CrystallineSymmetries}, we simply focus on momentum-independent, inversion-symmetric, $N_b=4$-band models,
\begin{equation}
    h_{\vec{k}} = \sum_{j=1,3}(a_j\sigma_0 + \vec{b}_j \cdot \vec{\sigma}) \otimes \tau_j,
\end{equation}
and choose $a_j$ and $(\vec{b}_j)_{1,2,3}$ uniformly in $[-1,1]$ and $[-1/3,1/3]$, respectively.

Apart from this explicit example, we stress once more that it is evident that the concept of considering topological sectors with respect to a reference state within a topological scheme of interest can in all generality be incorporated in our method, possibly involving the described procedure that allows for the closure of designated band gaps. Indeed, we note that even K-theory has a crucial intrinsic reference state, as it is by definition stable upon adding bands to the valence or conduction sector and band gaps within the filled or unoccupied sector can be freely closed without affecting the topology. 
\vspace{1em}

\section{Steepest ascent and symmetrization}
\label{SteepestAscent}
One might wonder whether the explicit symmetrization (\ref{ExplicitSymmetrization}) after obtaining $\varphi_{\vec{k}}$ from \equref{VarphiFormGD} can remove the property that the deformation $\ket{\psi_{n\vec{k}}^{l'}} \rightarrow U_{\vec{k}}\ket{\psi_{n\vec{k}}^{l'}}$ corresponds to gradient ascent of the similarity measure (\ref{RefinedSimilarityMeasure}). We show here, however, that this is not the case.

As a result of symmetrization, we effectively use the unitary transformation
\begin{equation}
    U_{\vec{k}} = e^{i \Lambda_{\vec{k}}}, \quad \Lambda_{\vec{k}} = \frac{1}{|\mathscr{G}|} \sum_{g\in\mathscr{G}} \varphi_{\mathcal{R}^{-1}_v(g)\vec{k}} \mathcal{R}^\pdagger_\psi(g) \Lambda  \mathcal{R}^\dagger_\psi(g).
\end{equation}
Inserting this into \equref{RefinedSimilarityMeasure}, we find that $\ket{\psi_{n\vec{k}}^{l'}} \rightarrow U_{\vec{k}}\ket{\psi_{n\vec{k}}^{l'}}$ leads to
\begin{widetext}
\begin{equation}
    S_{l,l'} \rightarrow S_{l,l'} - \frac{2}{N_k} \sum_{\vec{k}} \frac{1}{|\mathscr{G}|} \sum_{g\in\mathscr{G}} \varphi_{\mathcal{R}^{-1}_v(g)\vec{k}} \frac{1}{\widetilde{N}_b}\sum_{n=1}^{N_b} \sum_{n'\in \mathcal{S}_n} \text{Im}\left[\braket{\psi_{n\vec{k}}^l| \mathcal{R}^\pdagger_\psi(g) \Lambda  \mathcal{R}^\dagger_\psi(g)|\psi_{n'\vec{k}}^{l'}}\braket{\psi_{n'\vec{k}}^{l'}|\psi_{n\vec{k}}^{l}} \right] + \mathcal{O}(\varphi^2). \label{SymmetrizedChange}
\end{equation}
As discussed in the main text, the symmetries of the system allow to restrict the sum in \equref{RefinedSimilarityMeasure} and, thus, in \equref{SymmetrizedChange} to the fundamental domain of the Brillouin zone (or only a 1D irreducible path). Therefore, we can regard all the different $\varphi_{\mathcal{R}^{-1}_v(g)\vec{k}}$ as independent quantities in \equref{SymmetrizedChange} and the resulting gradient ascent expressions are
\begin{equation}
    \varphi_{\mathcal{R}^{-1}_v(g)\vec{k}} =  - \eta\frac{1}{\widetilde{N}_b}\sum_{n=1}^{N_b} \sum_{n'\in \mathcal{S}_n} \hspace{-0.3em} \text{Im}\left[\braket{\psi_{n\vec{k}}^l|\mathcal{R}^\pdagger_\psi(g) \Lambda  \mathcal{R}^\dagger_\psi(g)|\psi_{n'\vec{k}}^{l'}}\braket{\psi_{n'\vec{k}}^{l'}|\psi_{n\vec{k}}^{l}} \right], \quad \forall \, g\in\mathscr{G}, \label{TransformedGradientAscent}
\end{equation}
with learning rate $\eta$ [as in \equref{VarphiFormGD} of the main text].
Due to the fact that the Hamiltonian is by construction invariant under $\mathscr{G}$, $\mathcal{R}^\pdagger_\psi(g) h^l_{\mathcal{R}^{-1}_v(g)\vec{k}} \mathcal{R}^\dagger_\psi(g) = h^l_{\vec{k}}$, it holds $\mathcal{R}^\dagger_\psi(g)\ket{\psi^l_{n\vec{k}}}=e^{i\theta^l_{n\vec{k}}}\ket{\psi^l_{n\mathcal{R}^{-1}_v(g)\vec{k}}}$ and \equref{TransformedGradientAscent} just reduces to \equref{VarphiFormGD} with $\vec{k}$ in the full Brillouin zone. 
\end{widetext}

\end{appendix}

\end{document}